\documentclass[fleqn,twoside]{article}
\usepackage{espcrc2,cite,graphicx,amsmath,amsfonts,amssymb,mathrsfs}
\usepackage[figuresright]{rotating}

\newcommand{\ie}{{\it i.e.}}

\newcommand{\eg}{{\it e.g.}}

\newcommand{\cf}{{\it cf.}}

\newcommand{\eq}{Eq.}

\newcommand{\fig}{Figure}

\newcommand{\Ref}{Ref.}
\newcommand{\Refs}{Refs.}
\newcommand{\Sec}{Section}

\newcommand{\Tab}{Table}

\newcommand{\JHFSK}{\mbox{\sf T2K}}
\newcommand{\JHFHK}{\mbox{\sf T2HK}}
\newcommand{\NuFactII}{\mbox{\sf NuFact-II}}

\newcommand{\NUMI}{\mbox{\sf NO$\nu$A}}

\newcommand{\ReactorII}{\mbox{\sf Reactor-II}}

\newcommand{\stheta}{\sin^22\theta_{13}}
\newcommand{\deltacp}{\delta_\mathrm{CP}}
\newcommand{\ldm}{\Delta m_{31}^2}
\newcommand{\sdm}{\Delta m_{21}^2}
\newcommand{\equ}[1]{\eq~(\ref{equ:#1})}
\newcommand{\figu}[1]{\fig~\ref{fig:#1}}

\newcommand{\bi}{\begin{itemize}}
\newcommand{\ei}{\end{itemize}}

\newcommand{\capdef}{}
\newcommand{\mycaption}[2][\capdef]{\renewcommand{\capdef}{#2}%
       \caption[#1]{{\footnotesize #2}}}

\begin{document}

\title{
{\bf Direct test of the MSW effect by the solar appearance term
in beam experiments}}

\author{
{\large Walter Winter}\address[IAS]{{\it School of Natural Sciences,
Institute for Advanced Study, Einstein Drive, Princeton, NJ 08540, USA}}\thanks{E-mail:
{\tt winter@ias.edu}}}

\begin{abstract}
\noindent {\bf Abstract}
\vspace{2.5mm}

We discuss if one can verify the MSW effect in neutrino oscillations at a high confidence
level in long-baseline experiments.
We demonstrate that for long enough baselines
at neutrino factories, the matter effect sensitivity is, as opposed to the mass hierarchy sensitivity, not suppressed by $\stheta$ because it is driven by the solar oscillations
in the appearance probability. Furthermore, we show that for the parameter independent direct verification of the MSW effect at long-baseline experiments, a neutrino factory with a baseline of at least $6 \, 000 \, \mathrm{km}$ is needed. For superbeams, we do not find a $5\sigma$ discovery potential of the MSW effect independent of $\stheta$. We
finally summarize different methods to test the MSW effect.

\vspace*{0.2cm}
\end{abstract}

\maketitle

\section{Introduction}

It is now widely believed that neutrino oscillations are modified by matter effects, which is
often referred to as the Mikheev-Smirnov-Wolfenstein (MSW) effect~\cite{Mikheev:1985gs,Mikheev:1986wj,Wolfenstein:1978ue}.
In this effect, the coherent forward scattering
in matter by charged currents results in phase shifts in neutrino oscillations.
Since ordinary matter consists of electrons, but no muons or taus, the
$W$ boson exchange causes a
relative phase shift of the electron neutrino flavor. This relative phase shift
then translates into changes of the neutrino oscillation probabilities.

The establishment of the LMA (Large Mixing Angle) solution in solar neutrino oscillations
by the combined knowledge from SNO~\cite{Ahmad:2002ka}, KamLAND~\cite{Eguchi:2002dm},
and the other solar neutrino experiments has lead to ``indirect'' evidence for the MSW effect
within the sun. A more direct test of these matter effects would be the
``solar day-night effect'' (see \Ref~\cite{Blennow:2003xw} and references therein), where the solar neutrino flux can (during the night) be enhanced through matter effects in the Earth due to regeneration effects~\cite{Carlson:1986ui}.
So far, the solar day-night effect has not been
discovered at a high confidence level by Super-Kamiokande and SNO solar neutrino measurements~\cite{Fukuda:1998rq,deHolanda:2002pp}. A future very large water Cherenkov
detector used for proton decay ($\sim 7$  $\times$ Super-Kamiokande) could establish this
effect at the $4 \sigma$ confidence level within ten years~\cite{Bahcall:2004mz}.
 Similar tests could be performed with
supernova neutrinos~\cite{Lunardini:2001pb}, which, however, have a strong
(neutrino flux) model, detector position(s), and $\theta_{13}$ dependence~\cite{Dighe:2003jg}. In addition, strong matter effects can also occur in atmospheric neutrino oscillations
in the Earth~\cite{Akhmedov:1998ui,Petcov:1998su}. Since the muon neutrino disappearance
probability is, to first order in $\alpha \equiv \sdm/\ldm$ and $\sin
 \theta_{13}$, not affected by Earth matter effects~\cite{Akhmedov:2004ny}, testing the
matter effects in atmospheric neutrinos is very difficult. However,
the appearance signal of
future long-baseline experiments is supposed to be very sensitive towards matter effects in atmospheric neutrino oscillations (see, for example, \Refs~\cite{Krastev:1990gz,Freund:1999gy,Mocioiu:2000st,Ota:2000hf,Freund:2000ti}). This
makes the long-baseline test one natural candidate to directly discover the MSW effect at
a very high confidence level.

Since the direct verification of the MSW effect would be another consistency check for our picture of neutrino oscillations, we
study the potential of future long-baseline experiments to test matter versus vacuum oscillations. A similar measurement based upon matter
effects in neutrino oscillations is the mass hierarchy sensitivity, which assumes that
the matter effect is present and then tests the difference between the normal and inverted
mass hierarchies.
We will use this measurement in some cases for comparison in order to show the
similarities and differences to the matter effect sensitivity. Note that the
direct test of the MSW effect at neutrino factories was, for example, studied in \Ref~\cite{Freund:1999gy}. Since at that time the parameter
$\alpha \equiv \sdm/\ldm$ was very small
for the LMA best-fit values, the contributions from the solar terms in the appearance
probability were neglected and the MSW effect sensitivity was therefore
determined to be strongly suppressed by $\stheta$. We will show that the
now larger best-fit value of $\sdm$ (and thus $\alpha$) does not justify this assumption anymore.

\section{Analytical motivation and qualitative discussion}
\label{sec:analytical}

For long-baseline beam experiments, the electron or muon neutrino appearance probability $P_{\mathrm{app}}$ (one of the probabilities $P_{e \mu}$, $P_{\mu e}$, $P_{\bar{e} \bar{\mu}}$, $P_{\bar{\mu} \bar{e}}$) is very sensitive to matter effects, whereas the disappearance probability
$P_{\mu \mu}$ (or $P_{\bar{\mu} \bar{\mu}}$) is, to first order, not. The appearance probability can be expanded in the small hierarchy parameter $\alpha \equiv \Delta m_{21}^2/\Delta m_{31}^2$ and the small $\sin 2 \theta_{13}$ up to the second order as~\cite{Cervera:2000kp,Freund:2001pn,Akhmedov:2004ny}:
\begin{eqnarray}
P_{\mathrm{app}} & \simeq & \sin^2 2\theta_{13} \, \sin^2 \theta_{23} \, \frac{\sin^2[(1- \hat{A}){\Delta}]}{(1-\hat{A})^2}
\nonumber \\
&\pm&   \alpha  \sin 2\theta_{13} \, \sin \delta_{\mathrm{CP}} \,
 \sin({\Delta}) \, \xi(\hat{A},\Delta)
\nonumber  \\
&+&   \alpha  \sin 2\theta_{13} \,   \cos \delta_{\mathrm{CP}} \, \cos({\Delta}) \, \xi(\hat{A},\Delta)
 \nonumber  \\
&+&  \alpha^2 \, \cos^2 \theta_{23}  \, \sin^2 2\theta_{12} \, \frac{\sin^2(\hat{A}{\Delta})}{\hat{A}^2}. \nonumber \\
\label{equ:PROBMATTER}
\end{eqnarray}
Here $\Delta \equiv \Delta m_{31}^2 L/(4 E)$, $\xi(\hat{A},\Delta) = \sin 2\theta_{12} \cdot \sin 2\theta_{23} \cdot \sin(\hat{A}{\Delta})/\hat{A} \cdot \sin[(1-\hat{A}){\Delta}] / (1-\hat{A})$,
 and $\hat{A} \equiv \pm (2 \sqrt{2} G_F n_e E)/\Delta m_{31}^2$ with $G_F$ the Fermi coupling constant and $n_e$ the electron density in matter. The sign of the second term is positive for $\nu_{e} \rightarrow \nu_{\mu}$ or
$\nu_{\bar{\mu}} \rightarrow \nu_{\bar{e}}$ and negative for $\nu_{\mu} \rightarrow \nu_{e}$ or
$\nu_{\bar{e}} \rightarrow \nu_{\bar{\mu}}$. The sign of $\hat{A}$ is determined by the sign of $\Delta m_{31}^2$ and choosing neutrinos (plus) or antineutrinos (minus). Note that the matter effect in \equ{PROBMATTER} enters via the matter potential $\hat{A}$, where the
equation reduces to the vacuum case for $\hat{A} \rightarrow 0$ (\cf, \Ref~\cite{Akhmedov:2004ny}).

Since $\stheta>0$ has not yet been established, any suppression by $\stheta$ would be a major disadvantage for a measurement. Therefore, let us first investigate the interesting limit $\stheta \rightarrow 0$. In this limit, only the fourth term in \equ{PROBMATTER} survives, which is often referred to as the ``solar term'', since the appearance signal in the limit $\theta_{13} \rightarrow 0$ corresponds to the contribution from the solar neutrino oscillations. It would
vanish in the two-flavor limit (limit $\alpha \rightarrow 0$) and would grow proportional to $\left( \sdm L/(4E) \right)^2$ in vacuum (limit $\hat{A} \rightarrow 0$),
as one expects from the solar neutrino contribution in the atmospheric limit. Note that
this term is equal for the normal and inverted mass hierarchies, which means that it cannot be used for the mass hierarchy sensitivity. In order to show its effect
for the matter effect sensitivity compared to vacuum, we use $\Delta P \equiv  P_{\mathrm{app}}^{\mathrm{matter}} -  P_{\mathrm{app}}^{\mathrm{vac}} $.
We find from \equ{PROBMATTER}
\begin{eqnarray}
\Delta P^{\theta_{13} \rightarrow 0}
& \simeq & \alpha^2 \, \cos^2 \theta_{23} \,  \sin^2 2\theta_{12} \nonumber \\
& & \times \Delta^2 \left( \frac{\sin^2(\hat{A}{\Delta})}{\hat{A}^2 \Delta^2} - 1 \right) \, .
\label{equ:matter3}
\end{eqnarray}
Thus, this remaining effect does not depend on $\stheta$ and strongly increases with the baseline. In particular, the function $\sin^2(\hat{A}{\Delta})/(\hat{A}^2 \Delta^2)$ is maximal
(\ie, unity) for $\hat{A} \Delta \rightarrow 0$ and has its first root for $\hat{A}\Delta = \pi$ at the ``magic baseline'' $L \sim 7 \, 500 \, \mathrm{km}$.\footnote{At the magic baseline~\cite{Huber:2003ak}, the condition $\sin(\hat{A}{\Delta}) = 0$ makes all terms but the first in \equ{PROBMATTER} disappear
in order to allow a ``clean'' (degeneracy-free) measurement of $\stheta$. Note that the argument $\hat{A}\Delta$ evaluates to $\sqrt{2}/2 G_F n_e L$ independent of $E$ and $\ldm$, which means that it only depends on the baseline $L$. This also implies that the MSW effect in the
limit $\theta_{13} \rightarrow 0$ actually modifies the solar oscillation frequency, because the argument $\hat{A}\Delta$ in \equ{matter3} does not depend on $\ldm$.}
In the Earth, where \equ{PROBMATTER} is valid because of the
approximation $\Delta m_{21}^2 L/(4 E) \ll 1$,
we therefore have $\Delta P^{\theta_{13} \rightarrow 0} < 0$. This means that the matter effects will suppress the appearance probability, where maximal suppression is obtained at the magic baseline.  For short baselines, the expansion in
$\Delta$ shows that $\Delta P^{\theta_{13} \rightarrow 0} \propto L^4$
strongly grows with the baseline, and for very long baselines, the bracket in \equ{matter3} becomes close to $-1$, which means that $\Delta P^{\theta_{13} \rightarrow 0} \propto L^2$
compensated the $1/L^2$-dependence of the flux. Thus, we expect to be able to test the matter effect even for vanishing $\theta_{13}$ if the baseline is long enough.

There is, however, another important ingredient in these qualitative considerations: The statistics has to be good enough to detect the term suppressed by $\alpha^2$.
For the current best-fit values, $\alpha^2$ evaluates to $\sim 10^{-3}$.
One can easily estimate that the statistics of superbeams will normally be too low to measure the solar term for this value of $\alpha^2$ to a high accuracy: Let us compare the first and fourth terms in \equ{PROBMATTER}, which are suppressed by $\stheta$ and $\alpha^2$, respectively.
If one assumes that the other factors in the first and fourth terms are of order unity
(at least for $\Delta  \sim \pi/2$ close to the first oscillation maximum),
one can estimate for a specific experiment that the contribution from the $\alpha^2$-term
only becomes significant if the $\stheta$-sensitivity limit of this experiment
is much better than $\alpha^2$. This condition is,
in general, not satisfied for the proposed superbeams\footnote{In fact, for superbeams, the background from the intrinsic (beam) electron neutrinos limits the
performance, which means that increasing the luminosity would not solve this problem.} and could only be circumvented by a very long baseline, where the probability difference in \equ{matter3} grows $\propto L^2$.
For example, the \NUMI\ superbeam in the simulation of \Ref~\cite{Huber:2004ug} would only
lead to about four events with almost no dependence on the matter effect for $\theta_{13} \rightarrow 0$ (dominated by the intrinsic beam background). For neutrino factories,
however, this order of $\alpha^2$ should be accessible
for long enough baselines. For example, for the neutrino factory \NuFactII\ of \Ref~\cite{Huber:2002mx} at a baseline of $6 \, 000 \, \mathrm{km}$, we find for $\theta_{13} \rightarrow 0$ about 90 events in matter compared to 421 in vacuum.

Another interesting limit is the one of large values of $\stheta$, where
the first term in \equ{PROBMATTER} dominates, \ie, for $\sin 2 \theta_{13} \gg \alpha$, which
is equivalent to $\stheta \gg 10^{-3}$. It is strongly enhanced close to the matter resonance, where the resonance condition is given by $\hat{A} \rightarrow +1$. This condition evaluates to
a resonance energy of $\sim 9 \, \mathrm{GeV}$ (for $\rho = 3.5 \, \mathrm{g/cm^3}$ and $\ldm = 2.5 \cdot 10^{-3} \, \mathrm{eV}^2$) in the Earth's mantle, which is usually covered by a neutrino factory energy spectrum. Therefore,
neutrino factories are supposed to be very sensitive to matter effects, which are in this
limit driven by the atmospheric $\ldm$. Since the sign
of $\hat{A}$ depends on the mass hierarchy (and using neutrino or antineutrinos), it leads to
a strong enhancement (+) or suppression (-) of the appearance probability. One
therefore expects a very good sensitivity to the mass hierarchy for large enough $\stheta$ and
long enough baselines.
Similarly, one would expect a very good sensitivity to the matter
effect itself compared to the vacuum case, since the appearance probability in matter
becomes for long baselines very different from the vacuum case~\cite{Freund:1999gy}.
However, at least in the limit of small $\Delta$ or $\hat{A}$, the difference between the
matter and vacuum probabilities is by
about a factor of two smaller than the one between the normal and inverted hierarchy matter probabilities,
since the vacuum probability lies in between the other two. One can easily understand this
in terms of the matter potential which ``pulls'' the probabilities in two different directions apart from the vacuum case. In addition,
it is well known that the correlation with $\deltacp$ highly affects the mass hierarchy sensitivity in large regions of the parameters space (see, \eg, \Refs~\cite{Minakata:2001qm,Huber:2002mx}). Similarly, one can expect this correlation will destroy the matter effect sensitivity, too. Therefore, for large values of $\stheta$,
it is natural to assume that the test of the matter effect will be harder than the one of the mass hierarchy.

\section{Analysis methods and experiment simulation}

In general, we use a three-flavor analysis of neutrino oscillations, where we take into account
statistics, systematics, correlations, and degeneracies~\cite{Burguet-Castell:2001ez,Minakata:2001qm,Fogli:1996pv,Barger:2001yr}.
The analysis is performed with the $\Delta \chi^2$ method using the GLoBES software~\cite{Huber:2004ka}.

For the sensitivity to the matter effect, we test the hypothesis
of vacuum oscillations, \ie, we compute the simulated event rates for vacuum and a normal mass hierarchy. Note that there is not a large dependence on the mass hierarchy in vacuum, though the event rates depend (even in vacuum) somewhat on the mass hierarchy by the third term in \equ{PROBMATTER} (if one is far enough off the oscillation maximum). We then test this hypothesis
of vacuum oscillations by switching on the (constant) matter density profile and fit the
rates to the simulated ones using the $\Delta \chi^2$ method. In order to take into
account correlations, we marginalize over all the oscillation parameters and test both the
normal and inverted hierarchies. As a result, we obtain the minimum $\Delta \chi^2$
for the given set of true oscillation parameters which best fit the vacuum case.

For the mass hierarchy sensitivity, we compute the simulated rate vector for the chosen
mass hierarchy and fit it with the opposite sign of $\ldm$. Thus, it is  determined by the minimum $\Delta \chi^2$ at the $\mathrm{sgn}(\ldm)$-degeneracy~\cite{Minakata:2001qm}. Note that for neutrino factories
this minimum at the opposite sign of $\ldm$ might be very difficult to find because of
mixed degeneracies. Since we assume maximal mixing, the only relevant mixed degeneracy here is the $(\deltacp,\theta_{13})$-degeneracy~\cite{Burguet-Castell:2001ez} for the inverted $\ldm$.
The correlations originate in the minimization of the six-dimensional fit manifold at the
position of the $\mathrm{sgn}(\ldm)$-degeneracy, \ie, any solution with the opposite sign of
$\ldm$ fitting the original solution destroys the mass hierarchy sensitivity~\cite{Huber:2002mx}.
In addition, we assume a constant matter density profile with 5\% uncertainty,
which takes into account matter density uncertainties as well as matter profile
effects~\cite{Geller:2001ix,Ohlsson:2003ip,Pana}.

For all measurements, we assume that each experiment will provide the best measurement of the leading atmospheric oscillation parameters at that time, \ie, we use the information from the disappearance channels simultaneously. However, we have tested for this study that the disappearance channels do not significantly contribute to the matter effect sensitivity.\footnote{In fact, the disappearance channels alone could resolve the matter
effects for very large $L$ and large $\stheta$. However, in this region, the relative contribution of the disappearance $\Delta \chi^2$ to the total one is only at the percent level.} Furthermore, for the leading solar parameters, we take into account that the ongoing KamLAND experiment will improve the errors down to a level of about $10\%$ on each $\sdm$ and $\sin2\theta_{12}$~\cite{Gonzalez-Garcia:2001zy,Barger:2000hy}.

As experiments, we will mainly use neutrino factories based upon the representative \NuFactII\
 from \Ref~\cite{Huber:2002mx}.
In its standard configuration, it uses muons with an energy of $50 \, \mathrm{GeV}$,
$4 \, \mathrm{MW}$ target power ($5.3 \cdot 10^{20}$
useful muon decays per year), a baseline of
$3 \, 000 \, \mathrm{km}$, and a magnetized iron detector with a fiducial
mass of $50 \, \mathrm{kt}$. We choose a symmetric operation with $4 \, \mathrm{yr}$ in
each polarity. For the oscillation parameters, we use, if not stated otherwise, the current best-fit values $\ldm = 2.5 \cdot 10^{-3} \, \mathrm{eV}^2$,
$\sin^2 2 \theta_{23} = 1$, $\sdm = 8.2 \cdot 10^{-5} \, \mathrm{eV}^2$, and $\sin^2 2 \theta_{12} = 0.83$~\cite{Fogli:2003th,Bahcall:2004ut,Bandyopadhyay:2004da,Maltoni:2004ei}.  We only allow values for $\stheta$ below
the CHOOZ bound $\stheta \lesssim 0.1$~\cite{Apollonio:1999ae} and do not make any
special assumptions about $\deltacp$. However, we will show in some cases the results
for chosen selected values of $\deltacp$.

\section{Quantitative results}

\begin{figure}[t]
\begin{center}
\includegraphics[width=\columnwidth]{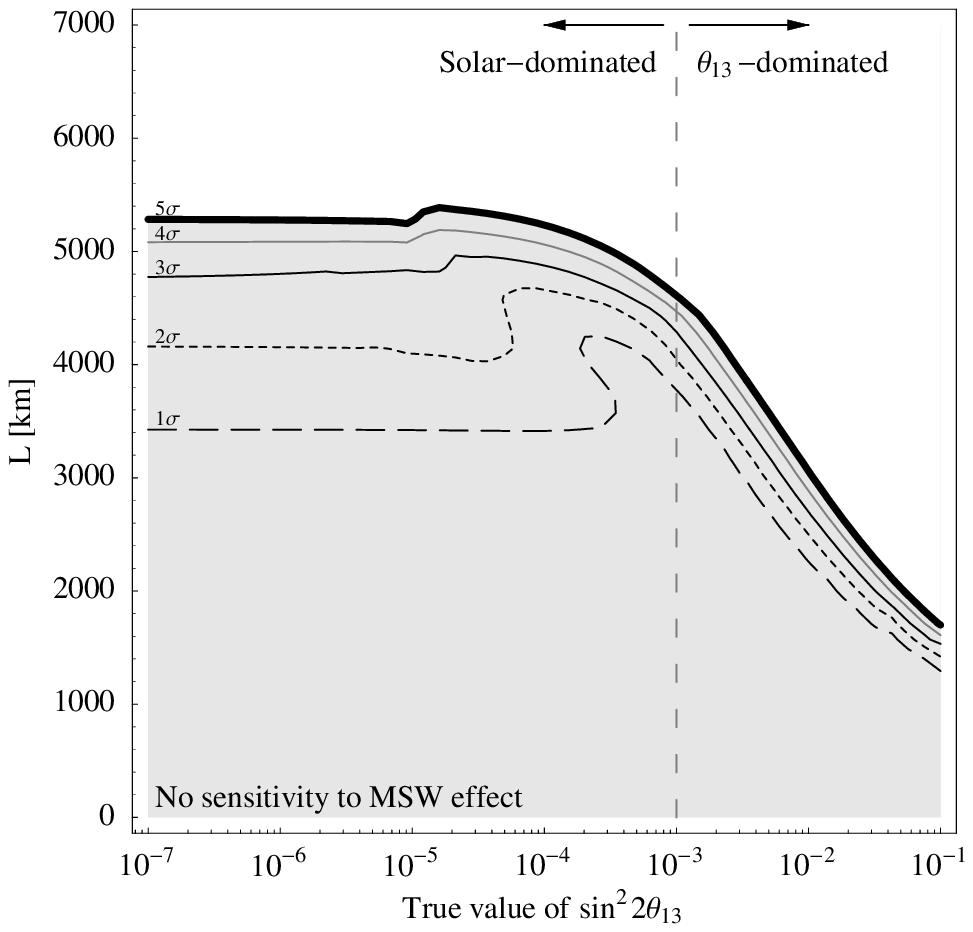}
\end{center}
\vspace{-1cm}
\mycaption{\label{fig:mswsens} Sensitivity to the MSW effect for \NuFactII\ as function
of the true value of $\stheta$ and the baseline $L$. For the simulated oscillation parameters,
the current best-fit values, $\deltacp=0$, and a normal mass hierarchy are assumed, whereas
the fit parameters are marginalized. Sensitivity is given at the shown confidence level on the
upper sides of the curves.}
\end{figure}

We show in \figu{mswsens} the sensitivity to the MSW effect for \NuFactII\
as function of the true values of $\stheta$ and the baseline $L$, where
$\deltacp=0$ and a normal mass hierarchy are assumed. The sensitivity
is given above the curves at the shown confidence levels. Obviously, the
experiment can verify the MSW effect for long enough baselines even for
$\stheta = 0$. The vertical dashed line separates the region where this
measurement is dominated by the first term ($\theta_{13}$-dominated) and the
fourth term (solar-dominated) in \equ{PROBMATTER}. It is drawn for $\stheta=10^{-3} \sim \alpha^2$, \ie, in this region all the terms of \equ{PROBMATTER} have similar magnitudes.
Obviously, the performance in the $\theta_{13}$-dominated (atmospheric oscillation-dominated) regime is much better than the one in the solar-dominated regime, because the $\theta_{13}$-terms provide information on the matter effects in addition to the solar term.
In this figure, the curves are shown for different selected confidence levels.
However, in order to really establish the effect, a minimum $5 \sigma$ signal will be
necessary. Therefore, we will only use the $5 \sigma$ curves below.

\begin{figure*}[t]
\begin{center}
\includegraphics[width=\textwidth]{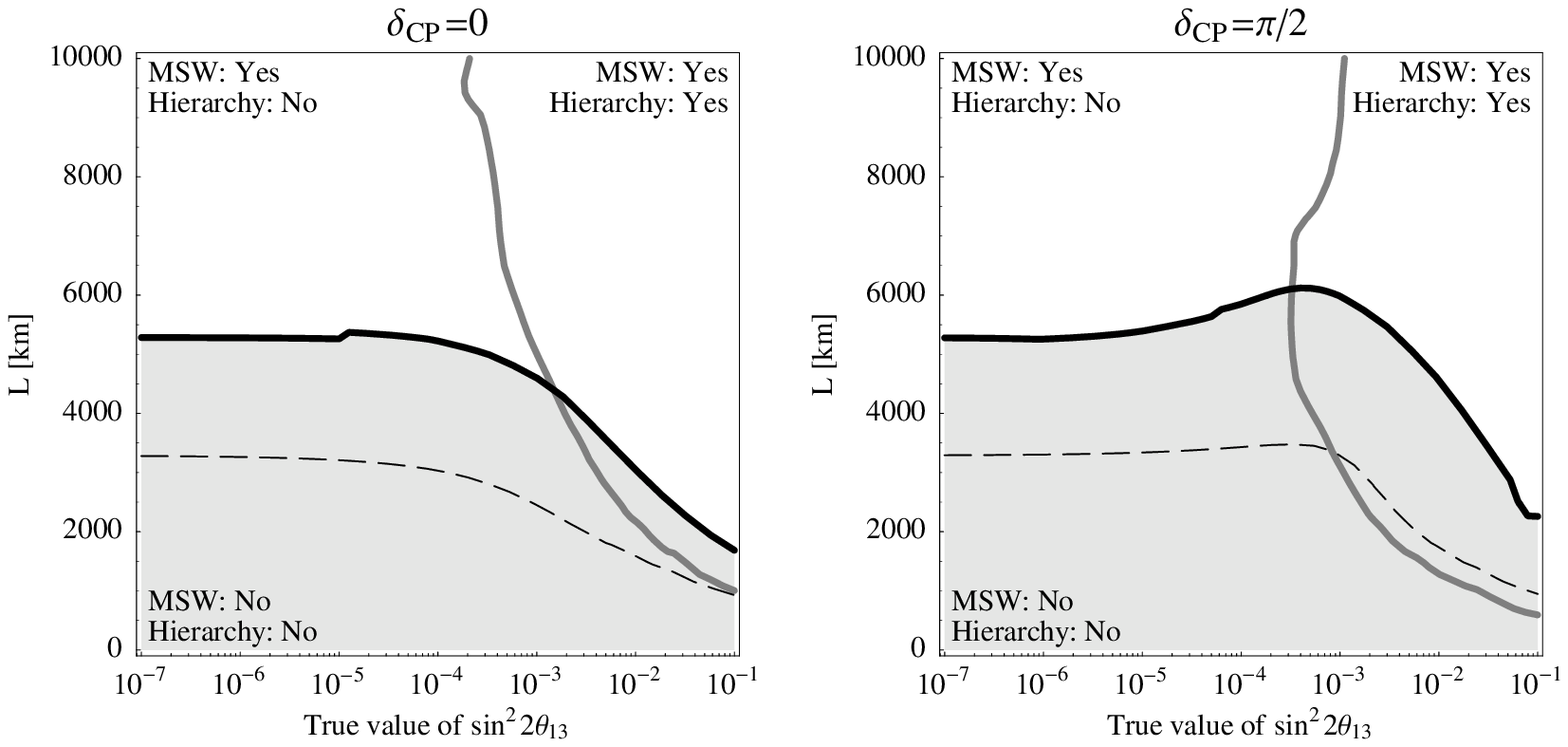}
\end{center}
\vspace{-1cm}
\mycaption{\label{fig:mswsign} The sensitivity to the MSW effect (black curves) and to
the mass hierarchy (gray curves) for \NuFactII\ as function
of the true value of $\stheta$ and the baseline $L$ ($5 \sigma$ only).
For the simulated oscillation parameters,
the current best-fit values, $\deltacp=0$ (left) or $\deltacp=\pi/2$ (right), and a normal mass hierarchy are assumed, wheras the fit parameters are marginalized over (solid curves).
Sensitivity to the respective quantity is given on the
upper/right side of the curves. The dashed curves correspond to the MSW effect sensitivity
without correlations, \ie, for all the fit parameters fixed.}
\end{figure*}

In order to discuss the most relevant parameter dependencies and to compare the
matter effect and mass hierarchy sensitivities, we show in \figu{mswsign} these
sensitivities for two different values of $\deltacp$. As we have tested, the true
value of $\deltacp$ is one of the major impact factors for these measurements. In addition,
the mass hierarchy sensitivity is modified by a similar amount for a simulated inverted instead of normal mass hierarchy, whereas the matter effect sensitivity does not
show this dependence (because the reference rate vector is computed for vacuum). As far as
the dependence on $\sdm$ is concerned, we have not found any signficant dependence of the
MSW effect sensitivity within the current allowed $3 \sigma$ range
 $7.4 \cdot 10^{-5} \, \mathrm{eV}^2 \lesssim \sdm \lesssim 9.2 \cdot 10^{-5} \, \mathrm{eV}^2$~\cite{Bahcall:2004ut}.
Hence, we show in \figu{mswsign} the selected two values of $\deltacp$
for estimates of the (true) parameter dependencies, since there are no major qualitative differences.

As one can see from this figure, the behavior of the MSW sensitivity for
short baselines and large $\stheta$ is qualitatively similar to the one of the mass hierarchy sensitivity, because both measurements are dominated by the $\theta_{13}$-terms of \equ{PROBMATTER}. However,
as we have indicated in \Sec~\ref{sec:analytical}, the difference between the normal and
inverted hierarchy matter rates is about a factor of two larger than the one between vacuum and matter rates (for any mass hierarchy). Thus, for large $\stheta$, the mass hierarchy sensitivity is better than the MSW sensitivity (better means that it works for shorter baselines). Note that the solar (fourth) term in \equ{PROBMATTER} is not dependent
on the mass hierarchy, which means that there is no mass hierarchy sensitivity for small
values of $\stheta$. In general, there are three regions for the MSW effect sensitivity
in \figu{mswsign}:
\begin{description}
\item[$\boldsymbol{\stheta \lesssim 10^{-5}}$:] Only the solar term in \equ{PROBMATTER}
is present. The MSW effect sensitivity therefore does not depend on $\deltacp$ or the
mass hierarchy.
\item[$\boldsymbol{\stheta \gtrsim 10^{-2}}$:] The measurement is dominated by the
first term in \equ{PROBMATTER} with some contribution of the second and third terms,
which means that there is some dependence on $\deltacp$.
\item[$\boldsymbol{10^{-5} \lesssim \stheta \lesssim 10^{-2}}$:] In the intermediate region,
these two effects are competing, which leads to the ``bump'' in the right panel of \figu{mswsign}. In particular, the relative contribution of the CP terms in \equ{PROBMATTER}
is quite large, which means that one expects the strongest $\deltacp$-dependence there.
\end{description}

For the MSW effect sensitivity, one can easily see from both panels of \figu{mswsign}
that for $\stheta \gtrsim 0.05$ a baseline of $3 \, 000 \, \mathrm{km}$ would be sufficient, because in this case the $\theta_{13}$-signal is
strong enough to provide information on the matter effects. However, in this case,
$\stheta$ will be discovered by a superbeam and it is unlikely that a neutrino factory
will be built. For smaller values $\theta_{13} < 0.01$, longer baselines will be necessary. In particular, to have sensitivity to the matter effect independent of the true parameter values, a neutrino factory baseline $L \gtrsim 6000 \, \mathrm{km}$ is a prerequisite. Therefore, this
matter effect test is another nice argument for at least one very long neutrino factory baseline.
Note that one can read off the impact of correlations with the oscillation parameters
from the comparison between the dashed and solid black curves in \figu{mswsign}.
If one just fixed all the oscillation parameters, one would obtain the dashed curves.
In this case, one could come to the conclusion that a shorter baseline would be sufficient,
which is not true for the complete marginalized analysis.

\begin{table*}[t]
\begin{center}
\begin{tabular}{lccp{6cm}}
\hline
Source/Method (where tested) & $\theta_{13}$-suppr. &  Reach [Ref.] & Comments/Assumptions \\
\hline
Solar $\nu$/Sun &  No & $6 \sigma$~\cite{Fogli:2004zn} & MSW effect in sun; by comparison
between vacuum and matter (existing solar $\nu$ experiments) \\
Solar $\nu$/Earth (``day-night'') &    No & $4\sigma$~\cite{Bahcall:2004mz} & By large Water Cherenkov detector used for proton decay \\
SN $\nu$/Earth, one detector &  No & n/a~\cite{Dighe:2003jg} & Observation as ``dips'' in
spectrum, but no observation guaranteed (because of flux uncertainties); effects depend on $\stheta$; HyperK-like detector needed \\
SN $\nu$/Earth, two detectors & No & $4\sigma -5 \sigma$~\cite{Lunardini:2001pb} & For SN distance $10 \, \mathrm{Kpc}$, $E_B = 3 \cdot 10^{53} \, \mathrm{ergs}$; at least two Super-K size detectors, depends on their positions \\
Atmospheric $\nu$/Earth & {\bf Yes} & $4\sigma$~\cite{Gandhi:2004bj} & Estimate for $100 \, \mathrm{kt}$ magn. iron detector computed for $\stheta=0.1$ \\
Superbeam/Earth $L \lesssim 5 \, 500 \, \mathrm{km}$ & {\bf Yes} & $2 \sigma$  & Estimate for \JHFHK -like setup for $\stheta \gtrsim 0.05$ at $L=3 \, 000 \, \mathrm{km}$; strongly depends on $\stheta$ and $\deltacp$  \\
Superbeam/Earth $L \gtrsim 5 \, 500 \, \mathrm{km}$ & No &  $\sim 3\sigma-4 \sigma$   & Estimate for \JHFHK -like setup independent of $\stheta$ \\
$\nu$-factory/Earth $L \lesssim 6 \, 000 \, \mathrm{km}$ &  {\bf Yes} &  $5 \sigma$   & Reach for $\stheta \gtrsim 0.05$ at $L=3 \, 000 \, \mathrm{km}$ ($\deltacp=\pi/2$); strongly depends on $\stheta$ and $\deltacp$ \\
$\nu$-factory/Earth $L \gtrsim 6 \, 000 \, \mathrm{km}$ &  No &  $5\sigma -8 \sigma$  & Range depending on $\deltacp$ for $L=6 \, 000 \, \mathrm{km}$;
for $L \gg 6 \, 000 \, \mathrm{km}$ much better reach, such as
$\sim 12 \sigma$ for $L = 7 \, 500 \, \mathrm{km}$ \\
\hline
\end{tabular}
\end{center}
\mycaption{\label{tab:summary} Different methods to test the MSW effect: Source and method (in which medium the MSW effect is tested), the suppression of the effect by
$\theta_{13}$, the potential confidence level reach (including reference, where applicable), and comments/assumptions which have led to this estimate.}
\end{table*}

As we have discussed in \Sec~\ref{sec:analytical}, the MSW test is very difficult for superbeams.
For the combination of \JHFSK , \NUMI , and \ReactorII\ from \Ref~\cite{Huber:2004ug}, it is not
even possible at the $90\%$ confidence level for $\stheta =0.1$ at the CHOOZ bound.
However, for a very large superbeam upgrade
at very long baselines, there would indeed be some sensitivity to the matter effect even for
vanishing $\theta_{13}$. For example, if one used  the \JHFHK\ setup from \Ref~\cite{Huber:2002mx} and (hypothetically) put the
detector to a longer baseline, one would have some matter effect sensitivity at the
$3 \sigma$ confidence level for selected baselines $L \gtrsim 5 \, 500 \, \mathrm{km}$. For the
``magic baseline'' $L \sim 7 \, 500 \, \mathrm{km}$, one could even have a $4 \sigma$ signal,
but $5 \sigma$ would hardly be possible.

\section{Summary and discussion}

We have investigated the potential of long-baseline experiments to test the matter
effect (MSW effect) in neutrino oscillations. In particular, we have discussed under
what conditions one can directly verify this MSW effect compared to vacuum oscillations
at a high confidence level.

Though it is generally known that beam experiments are, for sufficiently
long baselines, very sensitive to matter effects, we have demonstrated that the
$\theta_{13}$-terms in the appearance signal have much less matter effect sensitivity
than one may expect. Especially, the comparison with another matter effect-dominated
measurement, \ie, the mass hierarchy sensitivity, has shown that the MSW effect
sensitivity is much weaker for short baselines $L \lesssim 5 \, 000 \, \mathrm{km}$.
Note that both of these measurements suffer from correlations and degeneracies
especially for intermediate $\stheta$.

However, for long enough baselines $L \gtrsim 6 \, 000 \, \mathrm{km}$ and good
enough statistics, the solar term in the appearance probability is sensitive to
matter effects compared to vacuum, which means that the MSW effect sensitivity is not suppressed by $\stheta$ anymore.  Note that the solar term is not sensitive to the mass
hierarchy at all, but it is reduced in matter compared to vacuum. In summary, we have
demonstrated that a neutrino factory with a sufficiently long baseline would have good enough statistics for a $5 \sigma$ MSW effect discovery independent of $\stheta$, where the solar term becomes indeed statistically
accessible. However, a very long baseline superbeam upgrade, such as a
\JHFHK -like experiment at the ``magic baseline'' $L \sim 7 \,500 \, \mathrm{km}$,
could have some sensitivity to the solar appearance term at the $4 \sigma$ confidence level.

This result has three major implications: First, it is another argument for at least one
very long neutrino factory baseline, where the other purposes of such a baseline could be
a ``clean'' (correlation- and degeneracy-free) $\stheta$-measurement at the ``magic baseline''~\cite{Huber:2003ak} and a very good mass hierarchy sensitivity for large
enough $\stheta$. The verification of the MSW effect would be a little ``extra''
for such a baseline. In addition, note that
the mass hierarchy sensitivity assumes that the matter effects are present,
which means that some more evidence for the MSW effect
would increase the consistency of this picture.

Second, the absence of the $\stheta$-suppression in the solar
appearance term means that the direct MSW test at a beam experiment
could be competitive with others methods, for a summary, see  \Tab~\ref{tab:summary}.
However, it could be also partly complementary: If $\stheta$ turned out to be large,
it is the atmospheric oscillation frequency which would be modified by matter effects
and not the solar one.
In addition, the MSW effect in Earth matter could be a more ``direct'' test under
controllable conditions, because the Earth's mantle has been extensively studied by
seismic wave geophysics. Note that for atmospheric neutrinos, this test is much harder,
an example can be found in \Ref~\cite{Gandhi:2004bj}.

Third, we have demonstrated that the solar term in the appearance probability can really provide statistically significant information, which may also be useful for other applications.
For example, the dependence of the solar appearance term on $\cos \theta_{23}$ instead of $\sin^2 2 \theta_{23}$ in the disappearance probability could, for properly chosen
baselines, be useful to resolve the $(\theta_{23},\pi/2-\theta_{23})$-degeneracy~\cite{Diwan:2004bt}. Thus, the formerly
unwanted background term affecting any $\theta_{13}$ measurement could indeed be
useful for other applications.

\section*{Acknowledgments}

I would like to thank John Bahcall, Manfred Lindner,
and Carlos Pe$\tilde{\mathrm{n}}$a-Garay for useful
discussions and comments. This work has been supported by the W.~M.~Keck Foundation
and the NSF grant PHY-0070928.

\end{document}